\documentclass{emulateapj}
\usepackage{threeparttable}
\usepackage{bm}
\usepackage{chngpage}
\usepackage{graphicx}
\usepackage{mathrsfs}

\def\apjl{ApJL}

\shorttitle{Periodic FRBs as a Probe of EABs}
\shortauthors{Dai \& Zhong}

\begin{document}

\title{Periodic Fast Radio Bursts as a Probe of Extragalactic Asteroid Belts}

\author{Z. G. Dai\altaffilmark{1,2} and S. Q. Zhong\altaffilmark{1,2}}
\affil{\altaffilmark{1}School of Astronomy and Space Science, Nanjing University, Nanjing 210023, China; dzg@nju.edu.cn}
\affil{\altaffilmark{2}Key Laboratory of Modern Astronomy and Astrophysics (Nanjing University), Ministry of Education, Nanjing, China}

\begin{abstract}
The periodic activity of repeating fast radio burst (FRB) 180916.J0158+65 was recently reported by the CHIME/FRB Collaboration team. 28 bursts from this source not only show a $\sim16$-day period with an active phase of $\sim 4.0$ days but also exhibit a broken power law in differential energy distribution. In this paper, we suggest that FRB 180916.J0158+65-like periodic FRBs would provide a unique probe of extragalactic asteroid belts (EABs), based on our previously-proposed pulsar-EAB impact model, in which repeating FRBs arise from an old-aged, slowly-spinning, moderately-magnetized pulsar traveling through an EAB around another stellar-mass object. These two objects form a binary and thus the observed period is in fact the orbital period. We show that this model can be used to well interpret all of the observed data of FRB 180916.J0158+65. Furthermore, we constrain the EAB's physical properties and find that (1) the outer radius of the EAB is at least an order of magnitude smaller than that of its analogue in the solar system, (2) the differential size distribution of the EAB's asteroids at small diameters (large diameters) is shallower (steeper) than that of solar-system small objects, and (3) the two belts have a comparable mass.
\end{abstract}
\keywords{minor planets, asteroids: general -- pulsars: general -- radio continuum: general -- stars: neutron}

\section{Introduction}\label{introduction}
Since they were discovered for the first time \citep{Lorimer2007}, fast radio bursts (FRBs) have become one of the most mysterious astrophysical transients, because their physical origin remains unknown \citep{Petroff2019,Cordes2019,Katz2019,Platts2019}. Up to date, at least $100$ FRB sources have been detected, among which $\sim20$ sources show the repeating behavior (also see catalogue\footnote{http://www.frbcat.org}). The discovery of the first repeating source FRB 121102 \citep{Spitler2014} and the long-term follow-up observations \citep{Spitler2016,Scholz2016,Chatterjee2017,Marcote2017} indicate that all of the bursts from this source have a temporally-clustering feature,  providing an important clue to understanding an origin of FRBs.

Recently, the CHIME/FRB Collaboration team claimed to discover a periodically repeating source, FRB 180916.J0158+65, at $\nu\sim600$\,MHz \citep{CHIME2020}. This source is harbored in a massive spiral galaxy at redshift $z=0.0337\pm0.0002$ \citep{Marcote2020}, implying a luminosity distance $D_{\rm L}=149.0\pm0.9\,$Mpc for the Hubble constant $H_0=67.8\,{\rm km}\,{\rm s}^{-1}\,{\rm Mpc}^{-1}$. They detected 28 bursts from 16th September 2018 to 30th October 2019 and found a period of $16.35\pm0.18\,$days with an active phase of $\sim4.0\,$days \citep{CHIME2020}. The average burst rate is ${\cal{R}}_{\rm FRB}\sim 25\,{\rm yr}^{-1}$. In addition, the differential energy distribution of all of the bursts from this source reveals two power laws with indices of $-1.2\pm0.3$ and $-2.5\pm0.5$, connecting at a fluence $\sim 6.3\,{\rm Jy}\,{\rm ms}$ \citep[i.e., an isotropic-equivalent radio emission energy $\nu L_\nu\sim1.0\times10^{38}\,$erg,][]{CHIME2020}. A similar energy distribution can be seen for FRB 121102 with different radio telescopes \citep[for statistic analyses see][]{Gourdji2019,Wang2019a,Oostrum2019,Lin2020}. This shows that a turnover in the energy distribution of repeating FRBs seems to be ubiquitous, suggesting that it may be intrinsic.

Several models were proposed to explain the periodic activity of FRB 180916.J0158+65. In the first type of model, the $\sim16$-day period is due to magnetar free precession \citep{Levin2020,Zanazzi2020} or orbit-induced spin precession \citep{Yang2020} or fallback disk-induced precession \citep{Tong2020}. The basis of these studies is the early suggestion that repeating FRBs could originate from the magnetic activity of a magnetar \citep{Popov2013,Lyubarsky2014,Katz2016,Murase2016,Kashiyama2017,Metzger2017,Kumar2017,Beloborodov2017,Metzger2019}. The second type of model argued that the observed period is attributed to a binary period but the bursts could result from the distorted magnetic field lines of a pulsar immersed in a strong stellar wind of a massive companion \citep{Ioka2020}, following the cosmic combing model \citep{Zhang2017,Zhang2018}. A similar binary system scenario with a different bursting mechanism was proposed by \cite{Lyutikov2020} and \cite{Gu2020}. All of the works didn't discuss an energy distribution of the repeating bursts from FRB 180916.J0158+65 within the frame of a pulsar.

In this Letter, we suggest that FRB 180916.J0158+65-like periodic FRBs would provide a unique probe of extragalactic asteroid belts (EABs). Debris discs including asteroidal objects and their belts are widely thought to be the remains of the planet formation process. This is currently one of the {\em most interesting} topics in astronomy. The motivation of our study is based on the model of \cite{Dai2016}, in which repeating FRBs originate from an old-aged, slowly-spinning, moderately-magnetized pulsar traveling through an EAB around another stellar-mass object (possibly, a star or a white dwarf or a neutron star). Interestingly, if the two objects form a binary, then temporally clustering and even periodically repeating bursts would be naturally expected in this model, as discussed in \cite{Dai2016} and \cite{Bagchi2017} for FRB 121102. Furthermore, based on this model, observable radio bursts in the Milky Way galaxy were also predicted to arise from collisions between neutron stars and interstellar asteroids \citep{Siraj2019}. The remaining part of this paper is organized as follows. In section \ref{model} we constrain the physical properties (outer radius, mass and asteroidal size distribution) of an EAB by using the observed data of FRB 180916.J0158+65. We present discussion and conclusions in sections \ref{dis} and \ref{con}, respectively.

\section{Constraints on an EAB}\label{model}
\begin{figure}
\includegraphics[width=0.49\textwidth, angle=0]{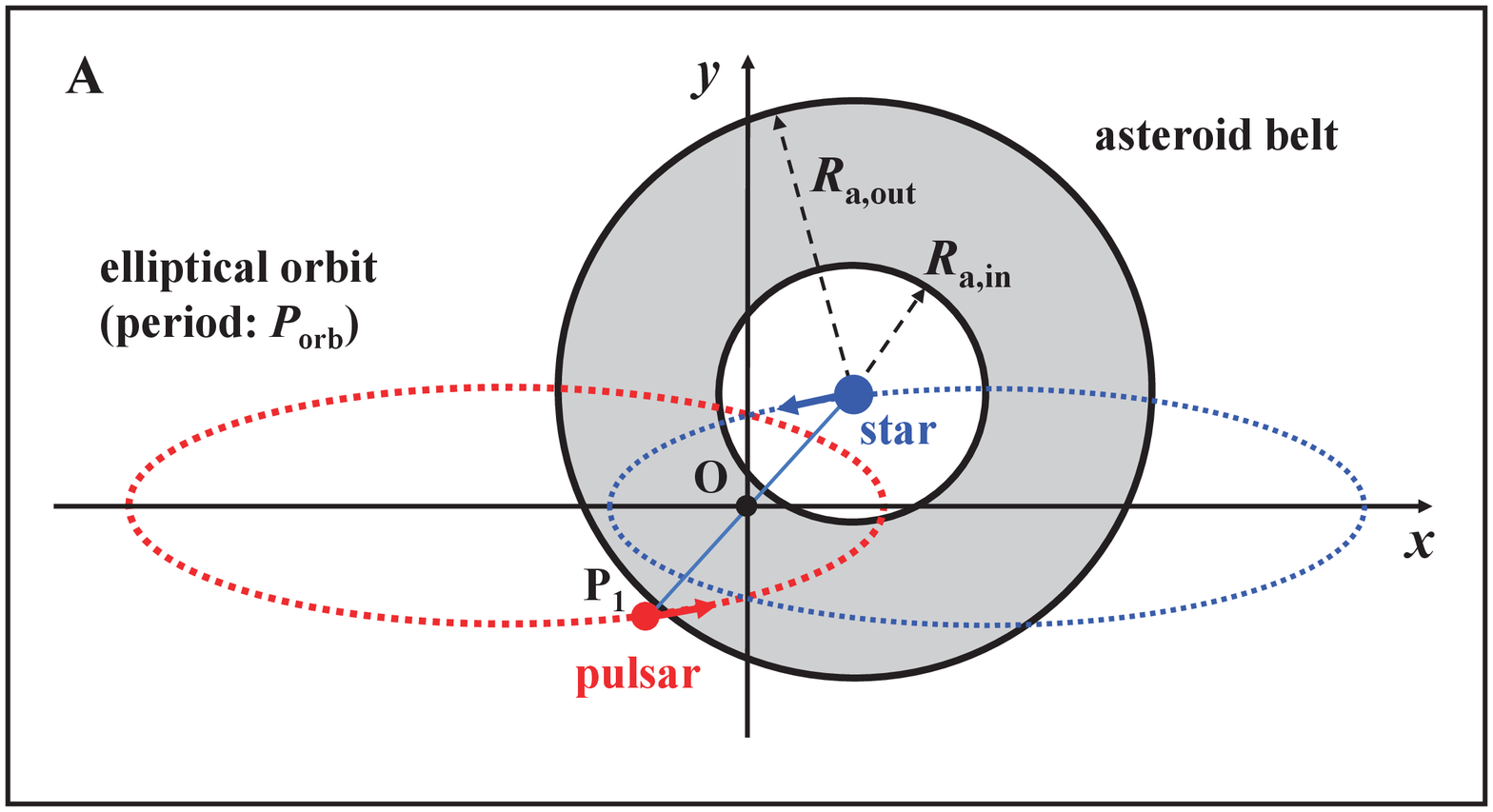}
\includegraphics[width=0.49\textwidth, angle=0]{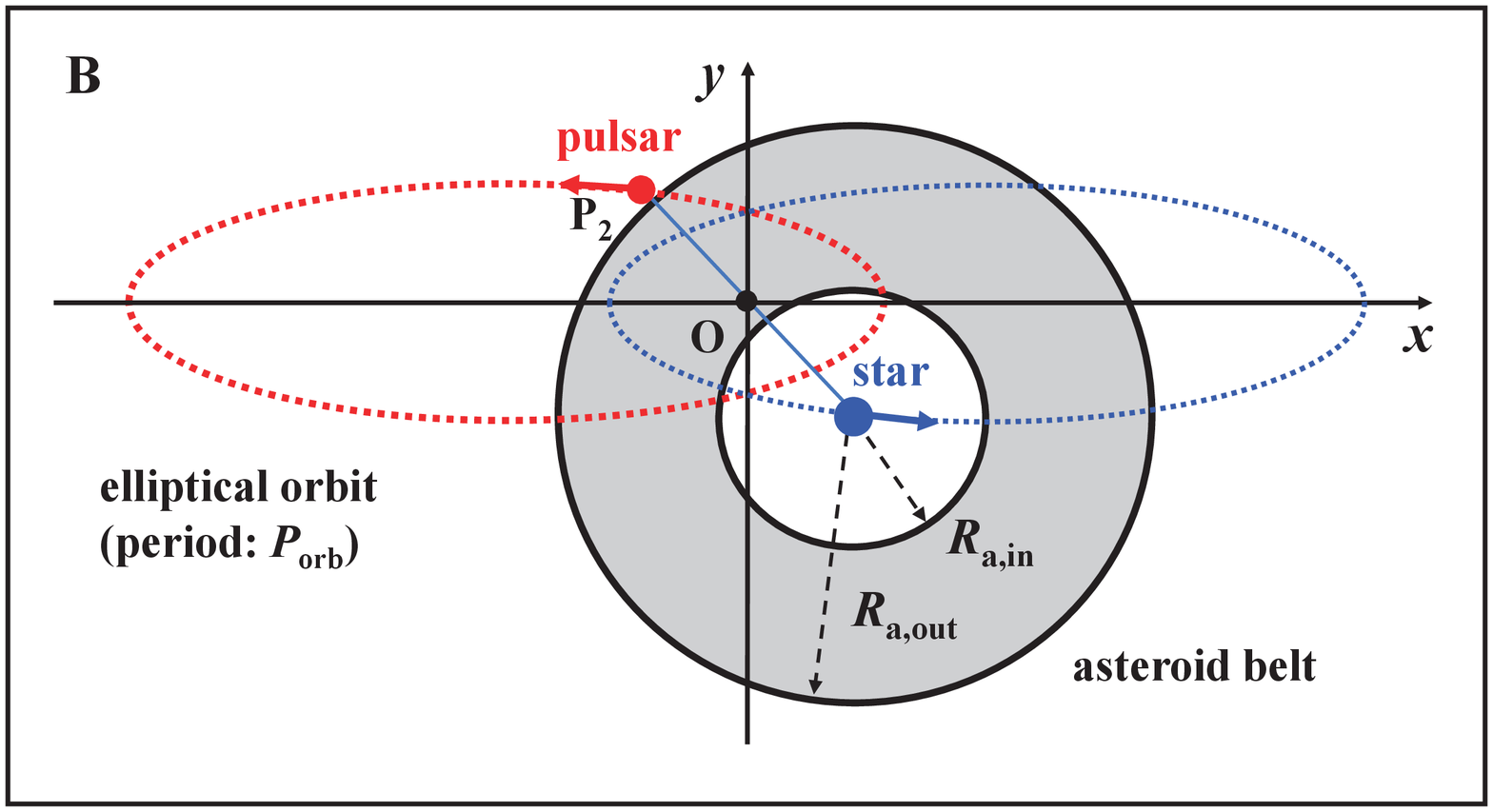}
\caption{Schematic picture of pulsar-EAB collisions. An old pulsar and a star with an EAB form a binary and rotate around their center of mass (point O), which is taken to be the original point of a coordinate system ($x,\,y$). The two objects move along respective elliptical orbits with an orbital period $P_{\rm orb}$. These orbits are assumed to be coplanar with the belt in order that pulsar-asteroid collisions are the most frequent. The pulsar first arrives at point P$_1$, at which it exactly enters the belt ({\em panel A}), and subsequently the pulsar reaches point P$_2$, at which it is just leaving from the EAB ({\em panel B}). The inner radius of the EAB is $R_{\rm a,in}$.}
\label{fig1}
\end{figure}

Following \cite{Dai2016}, we assume that a slowly-spinning ($P_{\rm pulsar}\gtrsim1\,$s), moderately-magnetized, wandering pulsar with an age $t_{\rm pulsar}\gtrsim 10^7$\,yr is captured by another stellar-mass object with a disc-shaped EAB of an outer radius $R_{\rm a,out}$. This EAB has an inner radius $R_{\rm a,in}$ and an orbital inclination angle\footnote{In the solar system, the main asteroid belt's inner radius $R_{\rm a,in}\sim 2.0\,$AU, outer radius $R_{\rm a,out}\sim 3.3\,$AU, and orbital inclination angle $\theta_{\rm a,incl}\sim 20^{\rm o}$ \citep{DeMeo2014}, so that the belt's thickness factor $\eta_{\rm t}=2\times\sin\theta_{\rm a,incl}\sim 0.7$ and width factor $\eta_{\rm w}=(R_{\rm a,out}-R_{\rm a,in})/R_{\rm a,out}\sim0.4$.}, implying that its thickness is nearly proportional to radius. In structure, the EAB may thus be analogous to the main asteroid belt in the solar system \citep{DeMeo2014,Pena2020} but the two belts could have some different physical parameters. The pulsar and the star, whose masses are taken to be $M_{\rm pulsar}$ and $M_{\rm star}$ respectively, form a binary (see Figure \ref{fig1}) and rotate around the center of mass (i.e., point O), which is also assumed to be the original point of a coordinate system ($x,\,y$). The two objects move along respective elliptical orbits with a period $P_{\rm orb}$. In the following, we investigate some constraints on the physical properties of the EAB by using the observed data of FRB 180916.J0158+65.

\subsection{Constraint on the Outer Radius}\label{radius}
In order to make pulsar-asteroid collisions the most frequent, we here consider a simple case in which the pulsar's elliptical orbit and the EAB are coplanar\footnote{Please see the second discussion on the probability of this case in section \ref{dis}.}. The lengths of the semi-major and semi-minor axes of the pulsar's elliptical orbit are $a$ and $b$, respectively, which are related with an orbital eccentricity through $e=(a^2-b^2)^{1/2}/a$. For FRB 180916.J0158+65, from Kepler's third law, the length $a$ for the {\em pulsar} is given by
\begin{eqnarray}
a & = & [G(M_{\rm pulsar}+M_{\rm star})]^{1/3}\left(\frac{P_{\rm orb}}{2\pi}\right)^{2/3}\left(\frac{1}{1+q}\right)\nonumber\\
& = & 2.7\times 10^{12}(1+q)^{-2/3}{\hat M}_{\rm star}^{1/3}{\hat P}_{\rm orb}^{2/3}\,{\rm cm},\label{period}
\end{eqnarray}
where $q\equiv M_{\rm pulsar}/M_{\rm star}$ is the mass ratio of the two objects, ${\hat M}_{\rm star}=M_{\rm star}/1.4M_\odot$, and ${\hat P}_{\rm orb}=P_{\rm orb}/16.35\,{\rm days}$. The two elliptical orbits satisfy
\begin{equation}
\frac{(x+ea)^2}{a^2}+\frac{y^2}{b^2}=1,\label{psr1}
\end{equation}
and
\begin{equation}
\frac{(x-eqa)^2}{(qa)^2}+\frac{y^2}{(qb)^2}=1,\label{star1}
\end{equation}
which correspond to the pulsar and the star, respectively.

As shown in panel A of Figure \ref{fig1}, when the star is at point ($x_{\rm star},\,y_{\rm star}$) (where it is required that $x_{\rm star}>0$ and $y_{\rm star}>0$), the pulsar reaches point P$_1$, whose coordinates are ($-x_{\rm star}/q,\,-y_{\rm star}/q$), at which the pulsar happens to arrive at a circular outer boundary of the EAB. Since this outer boundary satisfies the following equation
\begin{equation}
(x-x_{\rm star})^2+(y-y_{\rm star})^2=R_{\rm a,out}^2,\label{belt}
\end{equation}
when the pulsar reaches point P$_1$ the coordinates of its position are found from
\begin{equation}
x_{\rm star}^2+y_{\rm star}^2=\left(\frac{q}{1+q}\right)^2R_{\rm a,out}^2,\label{psr2}
\end{equation}
and
\begin{equation}
\frac{(x_{\rm star}-eqa)^2}{(q a)^2}+\frac{y_{\rm star}^2}{(q b)^2}=1.\label{psr3}
\end{equation}
From equations (\ref{psr2}) and (\ref{psr3}), thus, we can obtain ($x_{\rm star},\,y_{\rm star}$) if three parameters $e$, $q$ and $R_{\rm a,out}$ are given. In addition, we can also see from panel B of Figure \ref{fig1} that when the star reaches point ($x_{\rm star},\,-y_{\rm star}$), the pulsar is just leaving from the EAB, at which time the coordinates of the pulsar's position become ($-x_{\rm star}/q,\,y_{\rm star}/q$), namely point P$_2$.

\begin{figure}
\includegraphics[width=0.51\textwidth, angle=0]{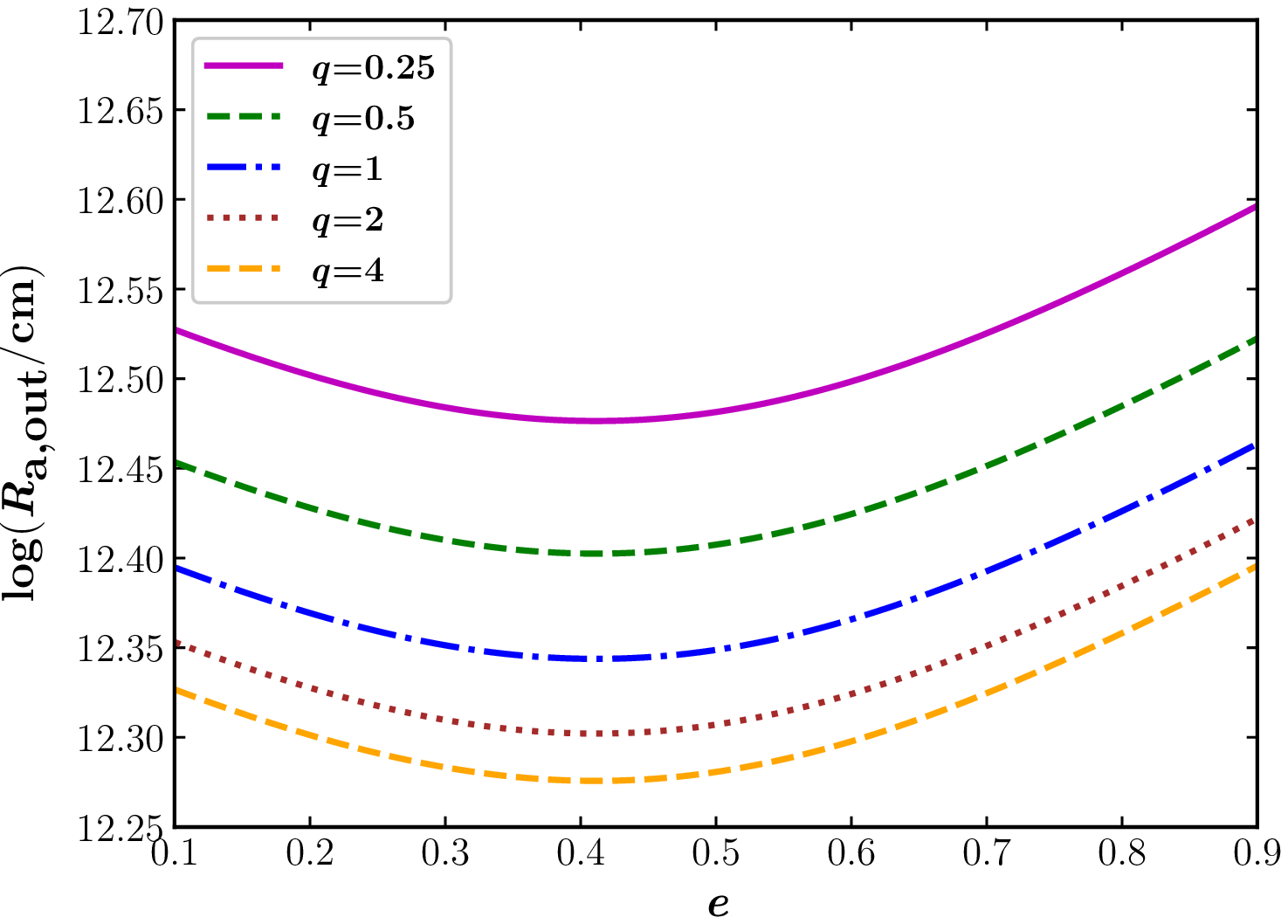}
\caption{$R_{\rm a,out}$ as a function of $e$ for $M_{\rm pulsar}=1.4M_\odot$ and $q=0.25,\, 0.5,\, 1,\, 2$, and $4$, in the case of FRB 180916.J0158+65 with an orbital period $P_{\rm orb}=16.35\,$days and a duty cycle $\zeta=0.24$ \citep[taken from][]{CHIME2020}.}
\label{fig2}
\end{figure}

The area swept out by a line between the pulsar and the center of mass from point P$_1$ to P$_2$ is calculated by
\begin{equation}
\Delta S_{\rm pulsar}=\frac{1}{2}\int_{-\theta_1}^{+\theta_2}r^2d\theta=\int_0^{\theta_2}\left[\frac{a(1-e^2)}{1+e\cos\theta}\right]^2d\theta,\label{Delta_S_p}
\end{equation}
where $\theta_1$ (or $\theta_2$) is the angle between the $x$-axis and the line OP$_1$ (or OP$_2$), $\theta_1=\theta_2=\pi-\arctan(y_{\rm star}/x_{\rm star})$. The total area enclosed by the pulsar's elliptical orbit is $S_{\rm pulsar}=\pi(1-e^2)^{1/2}a^2$. According to Kepler's second law, the ratio of these two areas is equal to the duration of the active time window ($\Delta P_{\rm orb}=4\,$days), in which the pulsar moves from point P$_1$ to P$_2$, divided by $P_{\rm orb}$. This means the duty cycle
\begin{equation}
\zeta\equiv\frac{\Delta S_{\rm pulsar}}{S_{\rm pulsar}}=\frac{\Delta P_{\rm orb}}{P_{\rm orb}}=\frac{4}{16.35}=0.24.\label{zeta}
\end{equation}
Therefore, under the condition of equation (\ref{zeta}), together with equations (\ref{psr2}), (\ref{psr3}), and (\ref{Delta_S_p}), we can numerically calculate $R_{\rm a,out}$ as a function of $e$ if the parameter $q$ is known. Figure \ref{fig2} shows $R_{\rm a,out}$ versus $e$ for $M_{\rm pulsar}=1.4M_\odot$ and five fixed values of $q$. We can see from this figure that $R_{\rm a,out}$ varies slowly with $e$ and has the minimum value at $e\sim 0.42$ for a given $q$. The outer radius increases from $\sim 0.13\,$AU to $\sim 0.26\,$AU if $q$ is set to be $0.25$ to $4$. This shows that $R_{\rm a, out}$ of the EAB responsible for FRB 180916.J0158+65 is at least an order of magnitude smaller than that of its solar-system analogue \citep{DeMeo2014,Pena2020}.

\subsection{Constraint on the Asteroidal Size Distribution}\label{size}
We consider an asteroid-pulsar collision. Following \cite{Colgate1981}, we assume that an asteroid as a solid body falls freely in the pulsar's gravitational field. This asteroid is originally approximated by a sphere with a mass $m$. It will first be distorted tidally by the pulsar at some breakup radius and subsequently elongated in the radial direction and compressed in the transverse direction. The timescale of such a bar-shaped asteroid accreted on the pulsar's surface is estimated by $\Delta t\simeq1.6m_{18}^{4/9}\,{\rm ms}$, where $m_{18}=m/10^{18}\,{\rm g}$ \citep[see equation 2 of][]{Dai2016}. This timescale is not only independent of the pulsar's radius but also weakly dependent on the other parameters such as the pulsar's mass and the asteroidal tensile strength and original mass density, even though the asteroid is assumed to be mainly composed of iron-nickel nuclei. The average rate of gravitational energy release near the stellar surface during $\Delta t$ is approximated by $\dot{E}_{\rm G} \simeq GmM_{\rm pulsar}/(R_{\rm pulsar}\Delta t)= 1.2\times 10^{41}m_{18}^{5/9}\,{\rm erg}\,{\rm s}^{-1}$, where and hereafter $M_{\rm pulsar}=1.4M_\odot$ and the pulsar's radius $R_{\rm pulsar}=10^6\,$cm are adopted. These simple estimates of $\Delta t$ and $\dot{E}_{\rm G}$ are well consistent with the observations of FRBs. This is why asteroid-pulsar collisions have been proposed as an origin model of FRBs \citep{Geng2015,Dai2016}. We now discuss the asteroidal size distribution in two following ways.

\subsubsection{A Simple Way}\label{way1}
We assume that $\xi$ is the efficiency of converting gravitational energy to radio emission and $f=\Delta\Omega/(4\pi)$ is the beaming factor of the emission (where $\Delta\Omega$ is the corresponding solid angle), so the isotropic-equivalent energy of an FRB can be estimated by
\begin{equation}
E_{\rm iso} \simeq (\xi/f)\dot{E}_{\rm G}\Delta t=1.9\times 10^{38}(\xi/f)m_{18}\,{\rm erg}.\label{Er1}
\end{equation}
This linearly proportional relation can provide an energy distribution of FRBs if both $\xi$ and $f$ are constants.

As shown by the Sloan Digital Sky Survey (SDSS) data \citep{Ivezic2001,Davis2002}, the Subaru Main Belt Asteroid Survey (SMBAS) data \citep{Yoshida2007}, the {\em Spitzer} Space Telescope infrared data \citep{Ryan2015}, and the High cadence Transient Survey (HiTS) data \citep{Pena2020} of solar system objects, the differential size distribution of the EAB's asteroids can be assumed to be written as
\begin{equation}
\frac{dN}{dD}\propto D^{-\beta}\propto\left\{
       \begin{array}{ll}
         D^{-\beta_1}, & D < D_{\rm br},\\
         D^{-\beta_2}, & D\ge D_{\rm br},
        \end{array}\label{nd}
        \right.
\end{equation}
where $D$ is the asteroidal diameter. In the solar system main belt, $\beta_1\simeq 2.3$, $\beta_2\simeq 4.0$ and $D_{\rm br}\sim 6.0\,$km \citep{Ivezic2001,Davis2002,Yoshida2007,Ryan2015,Yoshida2019,Pena2020}. A similar size distribution for Jupiter Trojans and Hildas was recently shown by \cite{Yoshida2019}. These authors comparatively studied the size frequency distributions of Jupiter Trojans, Hildas and main belt asteroids, and suggested that some physical mechanisms including collisional evolution and/or Yarkovsky effect during planet migration at the early solar system could provide a clue to understanding the origin of equation (\ref{nd}). In the case of $E_{\rm iso}\propto m$, this equation leads to a differential energy distribution of radio bursts,
\begin{equation}
\frac{dN}{dE_{\rm iso}}\propto E_{\rm iso}^{-\alpha}\propto \left\{
       \begin{array}{ll}
         E_{\rm iso}^{-\alpha_1}, & E_{\rm iso} < E_{\rm br},\\
         E_{\rm iso}^{-\alpha_2}, & E_{\rm iso}\ge E_{\rm br},
        \end{array}\label{ne}
        \right.
\end{equation}
where $\alpha=(\beta+2)/3$ and the break energy $E_{\rm br}\sim 1.7\times 10^{38}(\xi/f)(D_{\rm br}/6\,{\rm km})^3\,$erg is derived from equation (\ref{Er1}).

For FRB 180916.J0158+65, $\alpha_1=(\beta_1+2)/3\simeq 1.2$, $\alpha_2=(\beta_2+2)/3\simeq 2.5$, and $E_{\rm br}\sim 1.0\times 10^{38}\,$erg \citep[calculated from Extended Data Figure 3 of][]{CHIME2020}. These data imply that $\beta_1\simeq 1.6$, $\beta_2\simeq 5.5$, and $D_{\rm br}\sim 5.0(\xi/f)^{-1/3}\,$km. Therefore, the differential size distribution of the EAB's asteroids at small diameters (large diameters) is shallower (steeper) than that of asteroidal objects in the solar system.

\subsubsection{A Realistic Way}\label{way2}
\cite{Dai2016} explored asteroid-pulsar impact and radiation physics in detail and found that during such an impact an electric field induced outside of the asteroid has such a strong component parallel to the stellar magnetic field that electrons are torn off the asteroidal surface and accelerated to ultra-relativistic energies instantaneously. Subsequent movement of these electrons along magnetic field lines will cause coherent curvature radiation. From equation (15) of \cite{Dai2016}, the isotropic-equivalent emission luminosity is estimated by
\begin{equation}
L_{\rm iso}\sim 2.6\times 10^{40}(f\rho_{\rm c,6})^{-1}m_{18}^{8/9}\mu_{30}^{3/2}\,{\rm erg}\,{\rm s}^{-1},\label{Lr}
\end{equation}
where the beaming factor $f$ is introduced, $\rho_{\rm c,6}$ is the curvature radius of a magnetic field line near the stellar surface in units of $10^6\,{\rm cm}$, $\mu_{\rm 30}$ is the pulsar's magnetic dipole moment in units of $10^{30}\,{\rm G}\,{\rm cm}^3$, and the other parameters are taken for an iron-nickel asteroid \citep[also refer to equation 1 of][]{Siraj2019}. Thus, the isotropic-equivalent energy of an FRB is estimated by
\begin{equation}
E_{\rm iso}\simeq L_{\rm iso}\Delta t\sim 4.1\times 10^{37}(f\rho_{\rm c,6})^{-1}m_{18}^{4/3}\mu_{30}^{3/2}\,{\rm erg}.\label{Er2}
\end{equation}
This equation leads to an energy distribution of FRBs being similar to equation (\ref{ne}) but $\alpha=(\beta+3)/4$ and $E_{\rm br}\sim 3.6\times 10^{37}(f\rho_{\rm c,6})^{-1}(D_{\rm br}/6\,{\rm km})^4\mu_{30}^{3/2}\,$erg.

As clarified in \cite{CHIME2020}, only the CHIME/FRB telescopes detected radio bursts along the direction of FRB 180916.J0158+65 (and meanwhile, the 100-m Effelsberg telescope didn't detected any burst). This implies that the typical emission frequency of an FRB from this source is $\nu\sim600$\,MHz, which requires
\begin{equation}
\mu_{30}^{3/2}\rho_{c,6}\sim 10\chi^3,\label{mu_rho}
\end{equation}
where $\chi\lesssim1$ is introduced by assuming that $\chi \gamma_{\rm max}$ is the typical Lorentz factor of ultra-relativistic electrons emitting the FRB. Equation (\ref{mu_rho}) is derived from the maximum Lorentz factor ($\gamma_{\rm max}$) and curvature radiation frequency ($\nu_{\rm curv}\sim 600\,$MHz) of electrons given by equations (12) and (14) of \cite{Dai2016}, respectively.

From equations (\ref{nd}) and (\ref{ne}), we can see that $\beta_1\simeq 1.8$, $\beta_2\simeq 7.0$, and $D_{\rm br}\sim 7.6f^{1/4}\rho_{\rm c,6}^{1/4}\mu_{30}^{-3/8}\,{\rm km}\sim5.7f^{1/4}\chi_{-0.5}^{3/4}\mu_{30}^{-3/4}\,{\rm km}$, where $\chi_{-0.5}=\chi/10^{-0.5}$ and equation (\ref{mu_rho}) has been used. These results are basically consistent with the simple estimates in section \ref{way1}.

\subsection{Constraint on the Belt's Total Mass}\label{mass}
Since the geometric structure of the EAB may be somewhat similar to that of the solar-system main asteroid belt, we obtain the EAB's volume,
$V_{\rm belt}\sim 2\pi\eta_{\rm t}\eta_{\rm w}R_{\rm a,out}^3$, where $\eta_{\rm t}$ and $\eta_{\rm w}$ are assumed to be the EAB's thickness and width factors, respectively. If the asteroid-pulsar collision cross-section is taken to be $\sigma_{\rm a}$, from equation (18) of \cite{Dai2016}, the collision rate is given by
\begin{equation}
{\cal{R}}_{\rm a}\sim\frac{\sigma_{\rm a} v_{\rm pulsar}N_{\rm a}}{V_{\rm belt}},\label{R_a}
\end{equation}
where $N_{\rm a}$ is the total asteroid number in the EAB and hereafter $v_{\rm pulsar}\sim 10^7\,{\rm cm}\,{\rm s}^{-1}$ is the average velocity of the pulsar. Thus, the observed FRB rate reads ${\cal{R}}_{\rm FRB}\sim\zeta f{\cal{R}}_{\rm a}$ (where $\zeta=0.24$ is the duty cycle), that is,
\begin{equation}
{\cal{R}}_{\rm FRB}\sim 0.33N_{\rm a,6}f\left(\frac{\eta_{\rm t}\eta_{\rm w}}{0.25}\right)^{-1}\left(\frac{R_{\rm a,out}}{1\,{\rm AU}}\right)^{-3}\,{\rm yr}^{-1},\label{R_FRB}
\end{equation}
where $N_{\rm a,6}=N_{\rm a}/10^6$. For FRB 180916.J0158+65, from \cite{CHIME2020}, ${\cal{R}}_{\rm FRB}\sim 25\,{\rm yr}^{-1}$. Inserting this observed rate into equation (\ref{R_FRB}) gives $N_{\rm a}$. Therefore, the total mass of the EAB can be approximated by
\begin{eqnarray}
M_{\rm belt}\sim N_{\rm a}{\bar m} & \sim & 1.2\times 10^{-2}M_\oplus{\bar m}_{18}f^{-1}
\nonumber \\ & & \times\left(\frac{\eta_{\rm t}\eta_{\rm w}}{0.25}\right)\left(\frac{R_{\rm a,out}}{1\,{\rm AU}}\right)^3,\label{M_b}
\end{eqnarray}
where ${\bar m}={\bar m}_{18}\times10^{18}\,{\rm g}$ is the average asteroidal mass. As shown in Figure \ref{fig2}, $R_{\rm a,out}$ is $\sim0.13\,$AU to $\sim0.26$\,AU, so the EAB's total mass $M_{\rm belt}$ is in the range of $\sim2.6\times 10^{-5}M_\oplus{\bar m}_{18}f^{-1}(\eta_{\rm t}\eta_{\rm w}/0.25)$ to $\sim2.1\times 10^{-4}M_\oplus{\bar m}_{18}f^{-1}(\eta_{\rm t}\eta_{\rm w}/0.25)$. This mass is not only about four to five orders of magnitude smaller than that of the EAB inferred from the first repeating FRB 121102 \citep{Dai2016} but also comparable to the mass of the main asteroid belt in the solar system \citep[$\sim 5\times 10^{-4}M_\oplus$,][]{Krasinsky2002,Li2019}.

\begin{figure}
\includegraphics[width=0.51\textwidth, angle=0]{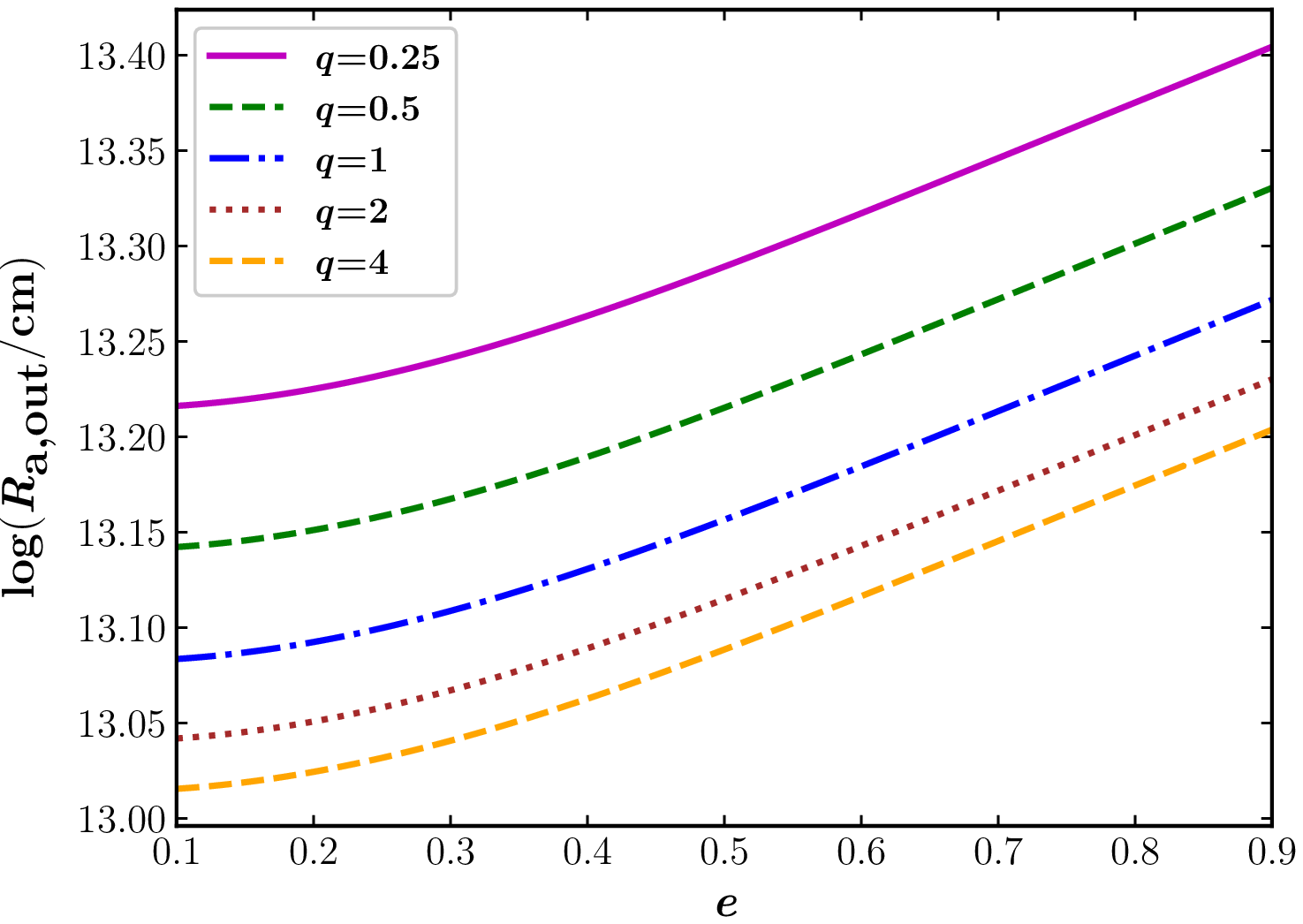}
\caption{$R_{\rm a,out}$ as a function of $e$ for $M_{\rm pulsar}=1.4M_\odot$ and $q=0.25,\, 0.5,\, 1,\, 2$, and $4$, in the case of FRB 121102 with an orbital period $P_{\rm orb}=159\,$days and a duty cycle $\zeta=0.47$ \citep[taken from][]{Rajwade2020}.}
\label{fig3}
\end{figure}

\section{Discussion}\label{dis}
We now discuss the validity and implications of our model. First, we have assumed an old-aged ($t_{\rm pulsar}\gtrsim10^7\,$yr), slowly spinning ($P_{\rm pulsar}\gtrsim1\,$s) pulsar, whose surface temperature cools as $T_{\rm s}\sim 6\times 10^4(t_{\rm pulsar}/10^7\,{\rm yr})^{-1}\,$K due to the fact that stellar surface black-body radiation becomes the dominant cooling mechanism \citep{Shapiro1983}. The resultant low cooling luminosity, together with an extremely low spin-down power, makes the effects of this pulsar on any asteroid entering its magnetosphere (i.e., evaporation and ionization) become insignificant \citep{Cordes2008}. Thus, the asteroid can be assumed to fall freely over the stellar surface.

Second, we have also assumed the coplanarity between the binary orbit and the EAB in section \ref{model}. This corresponds to the pulsar-EAB edge-on collision case of \cite{Dai2016}. The ratio of the rate for this case to total (edge-on plus head-on) collision rate can be estimated by $\eta_{\rm t}/2$, which is $\sim 0.35$ if the EAB, in structure, is analogous to the main asteroid belt in the solar system. Therefore, the probability of edge-on collisions may be comparable to that of head-on collisions.

Third, if a binary with a pulsar and a companion star at first arose from two stars and if an asteroid belt around the companion star was outside of the critical stable circular orbit fitted by \cite{Rabl1988} and \cite{Holman1999}, then this belt would be dynamically unstable and its orbit would be significantly changed during long-term evolution. In our model, fortunately, a wandering pulsar is captured by a star (also possibly a white dwarf or a neutron star) surrounded by an {\em extended} asteroid belt and then the two objects form a binary. In this case, numerical simulations of \cite{Smallwood2019} show that a Kuiper-belt-like outer region of the extended asteroid belt is observably distorted but a solar-main-belt-like inner region remains almost unchanged for a long integration time. Since $R_{\rm a,out}$ shown in Figure \ref{fig2} is in the innermost region of this belt, it would be expected that the EAB for FRB 180916.J0158+65 is dynamically quasi-stable for a period of time. How long is this time? Because only those asteroids swept out by the cross section of $\sigma_{\rm a}$ centering around the pulsar will be accreted onto the surface of the pulsar, from equation (\ref{R_a}), the typical lifetime of the binary system producing repeating radio bursts can be estimated by
\begin{eqnarray}
{\cal{T}}_{\rm FRB} & \sim & \frac{N_{\rm a}}{{\cal{R}}_{\rm a}}\sim \frac{V_{\rm belt}}{\sigma_{\rm a}v_{\rm pulsar}}\nonumber \\
& \sim & 0.72\times 10^6\left(\frac{\eta_{\rm t}\eta_{\rm w}}{0.25}\right)\left(\frac{R_{\rm a,out}}{1\,{\rm AU}}\right)^3\,{\rm yr}.\label{T_FRB}
\end{eqnarray}
This suggests that the lifetime ${\cal{T}}_{\rm FRB}$ is in the range of $\sim 1.6\times 10^3(\eta_{\rm t}\eta_{\rm w}/0.25)\,$yr to $\sim 1.3\times 10^4(\eta_{\rm t}\eta_{\rm w}/0.25)\,$yr for $R_{\rm a,out}$ given in Figure \ref{fig2} for FRB 180916.J0158+65.

Finally, the frequency down-drift in a burst was detected to occur for FRB 180916.J0158+65 \citep{CHIME2020}. Similarly to \cite{Wang2019} through an analysis of the movement of emitting bunches along magnetic field lines at different heights, our model can well explain the observed frequency down-drift rate and polarization \citep{Liu2020}.

\section{Conclusions}\label{con}
In this paper, we have suggested that periodic FRBs such as the recently-discovered periodic FRB 180916.J0158+65 would provide a unique probe of EABs, following the pulsar-asteroid belt impact model of \cite{Dai2016}, in which repeating FRBs originate from an old-aged, slowly-spinning, moderately-magnetized pulsar traveling through an EAB around a stellar-mass object (perhaps, a star or a white dwarf or a neutron star). It has been naturally expected that if the two objects form a binary, there should be temporally clustering and even periodically repeating bursts, as predicted in this model and implied by the early observations on the first repeating FRB 121102. We have shown that this model can be used to understand all of the observed data of FRB 180916.J0158+65, and provided some constraints on the EAB's physical properties. Our findings are as follows.
\begin{itemize}
\item The outer radius of the EAB responsible for FRB 180916.J0158+65 is at least an order of magnitude smaller than that of its solar-system analogue.
\item The power-law index of the differential size distribution of the EAB's asteroids at small diameters (large diameters) is smaller (larger) than the corresponding index of solar-system small objects.
\item The EAB's total mass is about four to five orders of magnitude smaller than that of the EAB inferred from the first repeating FRB 121102 and comparable to the mass of the main asteroid belt in the solar system.
\end{itemize}

{\em A Note Added}. After the submission of this paper, \cite{Rajwade2020} reported a periodicity search for FRB 121102 and found a tentative period of $159^{+3}_{-8}$ days in the periodogram with a duty cycle of $\sim47$\%. Interestingly, this result is well consistent with the possible periodic activity predicted by our model for FRB 121102 \citep{Dai2016,Bagchi2017}, and thus, from the analysis in this paper, can also provide a constraint on $R_{\rm a,out}$ of an EAB, shown in Figure \ref{fig3}. It is seen from this figure that $R_{\rm a,out}$ always increases with $e$ for a given $q$ and is in the range of $\sim0.69\,$AU to $\sim1.7\,$AU, which is smaller than that of the solar-system main asteroid belt by a factor of a few \citep{DeMeo2014,Pena2020}.

\acknowledgments
We would like to thank an anonymous referee for his/her helpful comments and suggestions that have allowed us to improve our manuscript, and thank Jonathan Katz, Dong-Zi Li, Jian Li, Fa-Yin Wang, Xue-Feng Wu, Yun-Wei Yu, Bing Zhang, Ji-Lin Zhou, and Li-Yong Zhou for their useful discussions. This work was supported by the National Key Research and Development Program of China (grant No. 2017YFA0402600) and the National Natural Science Foundation of China (grant No. 11833003).

\end{document}